\documentstyle[sprocl]{article}
\def\Journal#1#2#3#4{{#1} {\bf #2}, #3 (#4)}
\def\NPB{{\em Nucl. Phys.} B}
\def\PLB{{\em Phys. Lett.}  B}
\def\PRL{\em Phys. Rev. Lett.}
\def\PRD{{\em Phys. Rev.} D}

\def\UFN{\em Uspekhi Fiz. Nauk}

\def\SJNP{\em Sov. J. Nucl. Phys.}
\def\PISM{\em Pis'ma Z. Eksp. Teor. Fiz.}

\def\SJNP{\em Sov. J. Nucl. Phys.}
\def\ZETF{\em Zhurnal Teor. Exp. Fiz.}
\def\be{\begin{equation}}
\def\ee{\end{equation}}
\def\beq{\begin{equation}}
\def\eeq{\end{equation}}
\def\bea{\begin{eqnarray}}
\def\eea{\end{eqnarray}}

\def\bi{\bibitem}
\begin{document}
{\hbox to\hsize{July, 1997 \hfill TAC-1997-024}
\bigskip
\vglue .06in
\title{BARYOGENESIS, 30 YEARS AFTER.\\
\bigskip
 ({\it Lectures given at the 25th ITEP Winter School  }) }
\author{ A. D. DOLGOV  }
\address{Teoretisk Astrofysik Center, Juliane Maries Vej 30,\\
DK-2100, Copenhagen, Denmark
\footnote{Also: ITEP, Bol. Cheremushkinskaya 25, Moscow 113259, Russia.}
}

\maketitle\abstracts{
A review of the basic principles of baryogenesis is given. Baryogenesis in
heavy particle decays as well as electroweak, SUSY-condensate, and
spontaneous baryogenesis are discussed. The models of abundant creation
of antimatter in the universe are briefly reviewed.
 }

\section{Introduction}

This year is the anniversary of two great dates: 25th anniversary of ITEP
Winter
Schools and 30th anniversary of the seminal paper by A.D. Sakharov~\cite{ads}
on baryon asymmetry of the universe. In this paper a mechanism
was proposed, which explains the dominance of particles over antiparticles
in the universe by a dynamical evolution of an initially charge symmetric or
even arbitrary state. The necessary conditions for the generation of the
asymmetry, as formulated by Sakharov, are the following:
\begin{enumerate}
\item{}
Different interactions of particles and antiparticles, or in other words, a
violation of C and CP symmetries.
\item{}
Nonconservation of baryonic charge $B$.
\item{}
Deviation from thermal equilibrium in the early universe.
\end{enumerate}
We will discuss these conditions in more detail below and now return to the
history of the problem. The work by Sakharov remained unnoticed for several
years, except for a paper by Kuzmin \cite{kuz} in 1970, where a somewhat
different realization of the Sakharov's ideas was considered.
Still the hypothesis of
baryonic charge nonconservation, which is necessary for the generation of
an asymmetry between baryons and antibaryons, was not accepted by the
establishment till 1974, when the models of Grand Unification were put
forward\,~\cite{pt,gg}. After it was understood that nonconservation of baryons
might be quite a natural and general phenomenon, the attitude to the
possibility of a dynamical generation of baryon asymmetry of the universe has
been drastically changed and two almost simultaneous papers by
Ignatyev et al \cite{ikkt} and
by Yoshimura \cite{yosh} in 1978 stimulated
the flood of papers which remains unabated for 20 years. For the
review of the earlier stage of the theory of baryogenesis
one may address the papers \cite{dz,kt1}. Different scenarios
of baryogenesis at that early period were mostly based on the decay of heavy
particles in different versions of Grand Unification Models.
They all have a general feature that the characteristic energy scale for
the processes with B-nonconservation is extremely high, close to the Planck
scale, $M_{GUT} = 10^{16}$ GeV. Since then many other scenarios of baryogenesis
have been developed with baryon nonconservation at much lower energies,
sometimes almost "at hand". For a review of these models one can
address the paper \cite{ad1}. The most fashionable
for already several years remains electroweak baryogenesis \cite{krs}
where essential physical processes take place at the energy scale
around 100 GeV.
The dominant majority of papers are devoted to different realizations of
this particular model. Reviews of activity in this field can be found in
refs. \cite{ad1,ck1,rs}. By necessity they are all incomplete because new
papers on the subject constantly appear with a non-decreasing production rate.

In these lectures I will give an introduction to the theory of
baryogenesis for non-experts paying most attention to general features
of the phenomenon. I will not be able to cover very recent development
in the field. This would demand much more time, space, and skill. The content
of the lectures is the following. In the next four
Sections the general picture and the
three Sakharov's conditions are discussed. In Sec. 6  the baryogenesis through
decays of massive particles is considered. In Sec. 7 the other major
baryogenesis
scenarios are enumerated and three of them (electroweak, SUSY-condensate, and
spontaneous) are discussed in some detail. In the last section some models
leading to a large amount of antimatter are reviewed.

\section{Generalities.}

We know that for any known particle there exists the antiparticle with
exactly (within experimental accuracy) the same mass, $m = \bar m$ and
decay width, $ \Gamma = \bar \Gamma$, and opposite signs of all the
charges associated with this particle, $Q_j =-\bar Q_j$. In spite of this
striking symmetry which would naturally imply equal number densities
of particles and antiparticles in the universe, $n = \bar n$, the
observed picture is quite different: the universe (at least in our
neighborhood is predominantly populated by particles: protons,
neutrons, and electrons with a very small fraction of antiparticles
observed in cosmic rays, which all can be explained by secondary
origin in energetic particle collisions. It is not excluded of
course that the dominance of matter over antimatter is only local
and is realized inside a finite volume characterized by the linear
size $l_B$, while further away the picture is reversed,
and so on and so forth. In this way the universe could be
globally symmetric with respect to particles and antiparticles.
Even if this is true, it could be  achieved only with baryon nonconservation
because the separation of matter and antimatter on an astronomically large
scale \cite{omn} does not seem possible. The size of our matter domain
$l_B$ is known to be quite large,
roughly speaking $l_B > 10$ Mpc \cite{gst,stec} (in the recent paper~\cite{der}
much more restrictive bounds are presented). For a smaller $l_B$
one would expect a too large flux of energetic $\gamma$-rays coming
from the reaction of annihilation of $p\bar p$ into $\pi$-mesons
with the subsequent decay $\pi^0 \rightarrow 2\gamma$, which would
take place in the boundary area between the world and anti-world.
Another signature of domains of antimatter would be a distortion of
spectrum and isotropy of cosmic microwave radiation. The
value of $l_B$ permitted by such observations would be smaller if
matter and antimatter domains are separated by voids which may
appear because of an excessive pressure produced by the annihilation at
earlier stages of the evolution of the universe or because of low
density of matter and antimatter in the boundary regions, if the baryon
asymmetry changes sign gradually so that in the intermediate region the
asymmetry vanishes or is very small.

The convenient dimensionless number which characterizes the
magnitude of the baryon asymmetry of the universe is the ratio of
the baryonic charge density $(n_B - n_{\bar B})$ to the number
density of photons in cosmic microwave radiation,
\beq{
 \beta = { n_B - n_{ \bar B} \over n_\gamma} \approx 3\cdot 10^{-10}
\label{beta}
}\eeq
Here $n_\gamma = 411.4
(T_\gamma /2.736^o {\rm K})^3 {\rm  cm^{-3}}$ and $T_\gamma =
2.736^o {\rm K}$ is the temperature of the radiation.
Direct observations of baryonic matter in the universe give for $\beta$,
roughly speaking,  one third of the above number, in particular because a
large amount of baryons may be invisible (the recent discovery by the
Hubble Space Telescope of very faint galaxies gives a larger fraction of
directly observed baryonic matter). A more accurate estimate
can be obtained from primordial nucleosynthesis, from which the result
(\ref{beta}) is inferred. The abundances of
light elements (${^4 He},\, {^3 He},\, {^7 Li}$, and especially $^2 H$)
produced during the "first three minutes" are sensitive indicators of the
baryonic number density at that epoch (for the review given at an ITEP
Winter School see ref. \cite{ad2} and for a more up-to-date analysis see
e.g. the paper \cite{ns}).

We believe that the magnitude of $\beta $ at nucleosynthesis was the same
as it is now. If baryonic charge was conserved during the period from
the primordial nucleosynthesis till the present day, the
difference $(n_B - n_{\bar B})$ remains constant in the comoving volume,
i.e. in the volume which expands together with the universe, $V \sim a^3
(t)$, where $a(t)$ is the scale factor describing the universe expansion
(the separation, $l(t)$, of two distant objects changes with time as
$l(t)\sim a(t)l_0$). The number density of noninteracting photons also
remains constant in the comoving volume, $T_\gamma \sim 1/a(t)$ and
$n_\gamma \sim 1/a^3(t)$. In this approximation the ratio $\beta$ would
not change in the course of expansion. In fact $a^3 n_\gamma$ was not
constant in the early universe. Indeed the annihilation of massive
particles, when the temperature dropped bellow their masses, heated up
the primeval plasma and increased the photon number density. However
there is a conserved quantity, namely the entropy density,
\beq{
s= (p+\rho)/T = {2\pi^2 \over 45} g_s(T) T^3,
\label{s}
}\eeq
which stays constant in the comoving volume if thermal equilibrium is
maintained \cite{sw,kt2}. This is usually true with a very good accuracy
during most of the universe history. In equation (\ref{s})
$\rho$ and $p$ are respectively the energy and pressure densities
in the primeval plasma and the factor $g_s(T)$ is roughly speaking
equal to the number of particle species with $m<T$.
Because of that it is more convenient to introduce the quantity
$\beta_s = (n_B - n_{\bar B}) /s$ which does not change in the course of
expansion in a thermal equilibrium state. In the standard physical model
without new long-lived particles and unknown interactions, there is no
entropy production after $e^+e^-$-annihilation so that $n_\gamma a^3$
remains constant. Below $T=m_e$ this new quantity
$\beta_s$ practically coincides with $\beta$ while at higher temperatures
they may differ by one-two orders of magnitude due to contribution from
heavier particles. The original baryon asymmetry would be diluted by the
same amount. Another source of entropy and of the corresponding dilution
of the baryon asymmetry are possible phase
transitions in the early universe of the first and even of the second order.
They could also considerably diminish the previously generated asymmetry.

At higher temperatures when $T \geq m_B = O({\rm GeV})$ baryons and
antibaryons in the plasma were practically equally abundant $(n_B -
n_{\bar B} ) / (n_B + n_{\bar B}) \approx 10^{-9}$, while presently
$n_{\bar B} \ll n_B$ and this ratio is very close to 1. In other words the
universe which is 100\% asymmetric now was almost charge symmetric during
early stages, so we need to explain a very small breaking of this
symmetry. Still this very small breaking resulted in a huge amplification
of the present-day baryonic number density. In the case of a symmetric
universe baryons would efficiently disappear
at small temperatures due to mutual annihilation
and the number density of survived baryons can be evaluated
as \cite{zn,dz}
\beq{
n_B = n_{\bar B} \approx n_\gamma /( \sigma_{ann} m_B m_{Pl})
\approx 10^{-19} n_\gamma ,
\label{nb}
}\eeq
where $\sigma_{ann}$ is the cross-section of nucleon annihilation, $m_B
\approx 1 {\rm GeV}$ is their mass and $m_{Pl} = 1.2\cdot 10^{19} {\rm
GeV} $ is the Planck mass. So in a baryo-symmetric universe the number
density of baryons would be 9(!) orders of magnitude smaller than what is
observed in reality. If this were true then there would not be enough building
material for formation of celestial bodies and of course life would not
be possible in such almost empty universe. One may invoke the anthropic
principle for an explanation of the baryon asymmetry: in a symmetric
universe life would not be possible and only in an asymmetric world
there could be an observer to put the question why the universe is
asymmetric. It is interesting to estimate the magnitude of the asymmetry
which would allow formation of celestial bodies and ultimately
creation of life, but we
have a better option. The asymmetry can be generated dynamically and its
magnitude can be expressed through (possibly not yet known) fundamental
parameters of particle physics.

\section{Rise and Fall of Discrete Symmetries.}

Less than a half of a century ago (in fact before 1956) it was
firmly believed that physical laws are symmetric with respect to mirror
reflection, P, charge conjugation (transition from particles to
antiparticles), C, and time inversion, T. Though none of these symmetries
followed from any fundamental principle the belief was quite strong and
it was a great shock when it
was found that space parity is not conserved \cite{ly,wu} so that a
mirror reflected process could be physically impossible. It was assumed
immediately \cite{ldl} that simultaneous mirror reflection and charge
conjugation, CP, restore the symmetry so that for each process with particles
the mirror reflected process with antiparticles, and otherwise the same, is
possible. However the general attitude to the discrete symmetries was
changed towards "everything which is not forbidden is permitted" and the
search for CP-violation was discussed in the literature \cite{ior}.
In 1964 it was found that CP is also broken \cite{ccft} and after this
discovery life in the universe became possible (of course with
nonconservation of baryons).

The only discrete symmetry which survived to the present day is the
combination of all three transformations, CPT. It has a rather good
reason to exist: the so called CPT-theorem can be proven \cite{cpt}
which states that
any Lorents-invariant theory with positive definite energy and the
normal relation between spin and statistics is invariant with
respect to CPT-transformation. As a result of this symmetry masses of
particles and antiparticles and their {\bf total} decay widths must be
exactly equal. However the probabilities of specific channels should be
different for charged conjugated processes if both C and CP are broken.
One comment is in order here. In the lowest order in perturbation theory
the amplitude of the charged conjugated process is equal to the complex
conjugate of the original amplitude because of the hermicity of the
Lagrangian. Thus in this approximation the
probabilities of these processes are equal even if C and CP are violated.
To break this equality higher order corrections are necessary. These
corrections come from inelastic rescattering of the decay products. The
corresponding imaginary part can be calculated from the unitarity
condition for the scattering matrix, $SS^+ =1$. Introducing the amplitude
matrix in the usual way, $S=1+iT$, one gets
\beq{
i(T - T^+) = -TT^+
\label{t-t}
}\eeq
Thus there are two sources of imaginary parts in amplitudes: imaginary parts
in coupling constants related to C(CP)-nonconservation and dynamical imaginary
parts due to on-mass-shell rescattering corrections. Therefore the effects of
C(CP)-violation are always suppressed by an extra power of coupling constants.
It can be seen that elastic rescattering of the decay products produce
the same phase correction to particles and antiparticles and another
channel is necessary to get a nontrivial phase correction and to
break equality of the absolute values of
the amplitudes of charge conjugated processes.

Experimentally the first Sakharov's condition is well justified,
CP-violation is directly observed in the decays of $K^0$-mesons,
but it is not yet known what mechanism is responsible for it.
A knowledge of that is
very important for baryogenesis because baryogenesis took place at a much
larger energy scale where the data on the kaon decays cannot be simply applied.
Anyway it is known in principle that antiparticles are not just mirror
reflections of particles, they have essentially different interactions
and are produced with different probabilities in charge conjugated processes.

\section{Nonconservation of Baryonic Charge.}

In contrast to breaking of C and CP invariance, nonconservation of
baryonic charge has not yet been observed in direct experiments. The only
"experimental" evidence of baryonic charge nonconservation is the
existence of our baryo-asymmetric universe. A rather strong argument in
favor of nonconservation of baryons comes from the necessity of
inflation. We do not see today any alternative way to create our
smooth and flat universe
without a relatively long inflationary stage. During inflation the universe
expands in accordance with the law
\beq{
a(t) \sim  \exp ( Ht )
\label{eht}
}\eeq
where the Hubble parameter $H$ is (approximately) constant. The necessary
duration of the inflationary stage for the solution of the flatness and
horizon problems is about  60-70 Hubble times, $ Ht = 60-70$.
The Hubble parameter is related to the total energy density in the universe
as $ H = \sqrt{ 8\pi \rho_{tot} /3m_{Pl}^2} $.
To keep $H$ constant we need to have a constant energy
density which is naturally realized by a scalar field, the inflaton.

Let us assume now that baryonic charge is strictly conserved. The density
of baryonic charge at the present day with respect to the number density
of cosmic microwave photons is about $10^{-9} - 10^{-10}$
(see eq. (\ref{beta})).
It means that at higher temperatures the energy density associated with
baryonic charge with respect to the total energy density would be at least
that large and would remain more or less constant during the radiation
domination stage when the universe expanded as $a(t) \sim \sqrt {t}$.
Now if we go deeper back in time to the inflationary stage, the energy
density of matter would be in the form of an inflaton field and would remain
constant except for the energy density associated with the baryonic
charge, $\rho_B$. Since the charge is conserved, $\rho_B$ cannot be constant
and evolves as $\rho_B \sim
1/a^4 \sim \exp (-4Ht) $ for relativistic baryons or $\rho_B \sim 1/a^3
\sim \exp (-3Ht)$ for nonrelativistic baryons (the latter is rather
improbable). One can quite easily check that the total energy density
cannot remain (approximately) constant for more than 6-7 Hubble times.
It makes long inflation
impossible. Note that this is not simply the statement that to have the
observed baryon asymmetry today the preinflationary (initial) value of the
baryonic charge density should be unnaturally huge but that the presence
of conserved charge does not permit to have an approximately constant
energy density and correspondingly does not let
inflationary expansion last sufficiently long time.

Theoretically it is rather natural to expect that baryonic charge is not
conserved. Normally, as we understand it, charge conservation is
associated with a local (gauge) symmetry like $U(1)$ in electrodynamics.
Such a symmetry usually implies the existence of massless gauge bosons
("photons") which produce long-range forces. Existence of such
Coulomb-like forces would break the equivalence principle for elements
with a different mass-to-baryonic-charge ratio. No such violation has
been observed and this gives the upper limit \cite{lyb} for the coupling
of would-be baryonic photons to baryons, $\alpha_B < 10^{-44}$ (compare
with the electromagnetic coupling constant, $\alpha = 1/137$).
Such a smallness of the coupling probably means that there is no
conserved current related to baryonic charge. Moreover there are quite a
few theoretical models which predict that baryonic charge is indeed
nonconserved. To start with, there are grand unification theories which
put quarks and leptons on equal footing (into the same particle
multiplet). Thus there should be be transitions between quarks and leptons
which break both baryon and lepton number conservation.
This would give rise to proton decay or to neutron-antineutron
oscillations, unfortunately not yet discovered by experiment.
A plethora of supersymmetric models also predict baryonic charge
nonconservation (for a review see e.g. ref. \cite{miv}) which could occur at
the energies below the grand unification scale, $m_{GUT} = 10^{15} -
10^{16}$ GeV and potentially be more dangerous for proton decay.
Moreover the standard electroweak theory {\bf predicts} nonconservation of
baryonic charge through quantum corrections \cite{gh}. This
nonconservation is negligibly small at low energies but could be very much
enhanced at high temperatures comparable with the electroweak
scale \cite{krs} (see section 7).

Thus we can conclude that baryonic charge is most probably nonconserved.
Manifestation of its nonconservation are strongly suppressed at low
energies but at high energies or temperatures, which existed in the early
universe, the processes with a change in $B$ might be efficient and
produce an excess of baryons over antibaryons.

\section{Thermodynamics of Baryogenesis.}

For a gas or plasma in thermal equilibrium state with temperature $T$
the particle occupation numbers are given by the function:
\beq{
f(p) = {1 \over  \exp \left[ (\sqrt{p^2 + m^2}-\mu)/T \right] \pm 1} .
\label{fp}
}\eeq
The chemical potential $\mu$ is nonzero in equilibrium only if the
particles in question possess a conserved charge $Q$ and the
corresponding charge density is non-vanishing. Indeed the chemical
potentials of particles and antiparticles in equilibrium are related by
the condition $\mu + \bar \mu = 0$ and in the case that fermions bear the
charge $Q$ (in what follows we will take $Q=B$
and talk about baryon asymmetry), the charge density is given by
\bea
n_Q = n -\bar n   =
g_s\int{ d^3p /(2\pi)^3
 \over \exp \left[ (\sqrt{p^2 + m^2}-\mu )/T \right] +1}  \nonumber \\
- g_s\int { d^3p /(2\pi)^3
\over  \exp \left[ (\sqrt{p^2 + m^2}+\mu)/T \right]+1 }
\label{nq}
\eea
where $g_s$ is the number of spin states.
In the case of nonconserved charge the corresponding chemical potential
vanishes in equilibrium due to charge nonconserving reaction and thus
$n=\bar n$ and the asymmetry $n_Q$ is zero. Recall that in thermal
equilibrium the sum of chemical potentials of particles in the initial
state is equal to that in the final state, $\sum \mu_i = \sum \mu_f$.
Making the conclusion about the vanishing of $n_Q$
we have implicitly used CPT-theorem, by which masses of
particles and antiparticle are equal.  Otherwise there would be an
asymmetry even in equilibrium:
\beq{
n_B \approx {g_s q_B \over 4\pi^2} {(m^2 - \bar m^2 ) \over T^2} T^3
\int _1^\infty dy \, \exp (-my/T) \,\sqrt {y^2 -1},
\label{deltam}
}\eeq
where $q_B$ is the baryonic charge of particles in question.

Typically the rate of expansion is small in comparison with the reaction
rates. The former is given by the Hubble parameter which is related to
the total energy density in the universe by the Einstein equations:
\beq
H = \sqrt{ { 8\pi \rho \over 3m_{Pl}^2}} =
\sqrt{{8\pi^3 g_* \over 90}} {T^2 \over m_{Pl}}.
\label{h}
\eeq
Here we have substituted the equilibrium expression for the energy
density, $\rho = \pi^2 g_* T^4 /30$, where $g_*$ counts the number of
relativistic particle species in the primeval plasma, one for each bosonic
spin state and 7/8 for each fermionic spin state. Since $H$ is suppressed by
the inverse big Planck mass, the expansion is normally slower than the
reaction rates which are either $\Gamma_d \sim \alpha m$ for decays of
particles with mass $m$ or $\Gamma_r \sim \alpha^2 T $ for reactions.
Still equilibrium is always somewhat broken for massive particles. To
see this let us consider the kinetic equation in the expanding universe:
\beq
{df \over dt} = {\partial f \over \partial t } -
Hp {\partial f \over \partial p} = I [ f ] ,
\label{dfdt}
\eeq
where we used the relation $ dp/dt = -Hp $ (red shift of momenta in the
expanding universe). The r.h.s. of this equation is a collision integral
to be specified below. Here we need only to know that $I[f]$ is annihilated
by the equilibrium distributions (\ref{fp}). Substituting them into the
l.h.s. we get:
\beq
{\rm l.h.s.} = {\exp [(E -\mu)/T] \over
\left[ \exp \left( (E-\mu)/T \right) + 1 \right]^2 }
\left( {\mu \dot T \over T^2} -{\dot \mu \over T} - {E \dot T \over T^2} -
{H p^2 \over TE} \right) .
\label{lhs}
\eeq
This expression can vanish for arbitrary $p$ only if the
particles in question are massless, so that $p=E$, and if the
following conditions are fulfilled:
$\dot \mu / \mu = \dot T / T$ and $\dot T / T = - H$.
So we can estimate the deviation from equilibrium as
\beq
{\delta f / f_{eq}} \approx {\left( Hm^2 / TE\Gamma\right)},
\label{dff}
\eeq
where $\Gamma$ is the characteristic rate of the reactions given by the
collision integral in the r.h.s.
One can roughly estimate this ratio at $T\approx m$
as $\delta f /f \approx 10m /\alpha^k
m_{Pl}$, where $\alpha$ if the characteristic coupling constant (at the
grand unification scale $\alpha \approx 1/50$) $k = 1$ or 2,
depending upon whether decay or two-body reactions are essential, and we take
$g_* \approx 100$. The deviation from equilibrium are noticeable either for
very heavy particles with the mass around $10^{16} - 10^{15}$ GeV
or for very weakly coupled ones with the coupling constant much
smaller than $\alpha$. For electroweak interactions at the electroweak
energy scale, $T=O({\rm TeV})$, the ratio (\ref{dff}) is close to
$10^{-15}$.

Now let us have a closer look at the collision integral. It has the
following form:
\begin{eqnarray}
I[f] = \int \Pi' {d^3 p_i \over (2\pi)^3\,2E_i}
\Pi {d^3 p_f \over (2\pi)^3\,2E_f} (2\pi)^4
\delta \left( \sum p_i - \sum p_f \right)
\nonumber \\
\left[ |A_{if}|^2 \Pi f_i \Pi (f_f \pm 1)-
|A_{fi}|^2 \Pi f_f \Pi (f_i \pm 1) \right] ,
\label{if}
\end{eqnarray}
where the integration is made over all final momenta and all initial ones,
except for the particle under scrutiny, which distribution function
enters the l.h.s. of equation (\ref{dfdt}). It is usually assumed that
T-invariance is true so that the amplitudes of direct and
inverse processes are equal up to a trivial change of signs of
momenta, $|A_{if}|^2 = |A_{fi}|^2$. This is the very well known detailed
balance condition. One can check that if this condition
is true the collision integral vanishes on the equilibrium functions
(\ref{fp}). The product $\Pi f_i \Pi (f_f \pm 1) $ is equal to the
product with $i$ and $f$ interchanged
because of the conservation of energy, $\sum E_i =\sum
E_f$, and chemical potentials, $\sum \mu_i = \sum \mu_f$, in equilibrium.
We know however that CP-invariance is broken, so by
CPT-theorem T-invariance also cannot be true, and the detailed balance
condition is violated. One might worry if breaking of CP would
simultaneously break normal equilibrium statistics. This is not the case
however because, as we have already noted, CP- and correspondingly
T-violation can be observable only when there are several different
processes giving rise to inelastic rescattering in the final state. If
these additional (and necessary) processes are taken into account,
the total contribution of all of them ensures vanishing of the
collision integral even if detailed balance is violated.
Equilibrium statistics is more general than the detailed balance condition.
In fact the validity of the canonical equilibrium distributions
follows either from the unitarity of S-matrix or from CPT-theorem
and conservation of probability. One of those is sufficient to
maintain the so called cyclic balance (instead of the detailed one)
when a cycle of several processes ensures thermal equilibrium.
This problem is considered in detail in ref. \cite{ad3}.

\section{Baryogenesis in Massive Particle Decays.}

This is the simplest and historically first model of baryogenesis. It has
a natural theoretical frameworks of grand unified models. As we have
already mentioned baryonic charge is not conserved in these models and,
since the gauge and Higgs bosons in these theories have very large
masses, $m= 10^{15}-10^{16}$ GeV, one may expect that their number density
could be essentially above the equilibrium one so that their
B-nonconserving and
charge asymmetric decay could produce a noticeable asymmetry between
baryons and antibaryons. As a simple model let us assume that there is a
charge
symmetric collection of $X$ and $\bar X$ gauge bosons of grand unification,
$n_X = n_{\bar X}$, and no other particles.
These bosons are known to have the following decay modes:
\beq
X \rightarrow qq, \,\,\, X \rightarrow  q\bar l,
\label{xqq}
\eeq
\beq
\bar X \rightarrow \bar q \bar q, \,\,\,
X \rightarrow \bar q  l \, .
\label{xbarq}
\eeq
where $q$ and $l$ are correspondingly quarks and leptons.
It is clear that baryonic charge is not conserved in these reactions because
the same initial state decays into particles with different baryonic
charges. If C and CP are not conserved the widths of charge conjugated
decay channels may be different:
\bea
\Gamma_{X\rightarrow qq} = (1+\Delta_q) \Gamma_q, \,\,\,
\Gamma_{X\rightarrow  q \bar l} = (1-\Delta_l) \Gamma_l \, , \nonumber \\
%\label{gqq}
%\eeq
%\beq
\Gamma_{\bar X \rightarrow \bar q \bar q } = (1-\Delta_q) \Gamma_q, \,\,\,
\Gamma_{\bar X \rightarrow \bar q l} = (1+\Delta_l) \Gamma_l.
\label{gbarq}
\eea
Here $\Delta_{q,l}$ are nonzero due to breaking of charge symmetries. If
these decay modes are the only ones, then by CPT-theorem the total decay
widths of $X$ and $\bar X$ should be the same and so $\Delta_q \Gamma_q =
\Delta_l \Gamma_l$.

If the energy of
fermions produced by these decays quickly drops down due to expansion or
thermalization by rescattering, so that baryonic
charge becomes effectively conserved in their collisions, then the baryon
asymmetry would be roughly equal to
\beq
\beta \approx {{4\over 3} \Delta_q \Gamma_q -
{2\over 3} \Delta_l \Gamma_l \over \Gamma_{tot} }\, { n_X \over n_0} ,
\label{betagut}
\eeq
where $n_X$ is the number density of the initial $X$-bosons and $n_0$ is
the number (or entropy) density of the produced light particles. The
latter may be considerably larger than just $2n_X$ because of a possible
increase of the number density of light particles
in the process of thermalization. Neglecting the universe expansion we can
estimate the number density of light particles as
$n_0 \approx  \rho_X ^{3/4} \approx n_X(m_X^{3/4}/ n_X ^{1/4})$
(the last equality is true for nonrelativistic X-bosons).

If the processes of thermalization and of the cooling by expansion are not
fast enough, the produced quarks and leptons would be quite energetic so
that baryonic charge would be nonconserved in their reactions. Such
reactions could wash out the baryon asymmetry produced by
the decays of X-bosons. It is
often stated in the literature that the asymmetry is erased by inverse
decays, $ qq \rightarrow X$, $ q \bar l \rightarrow X$, etc. However one
can see that this is incorrect. Indeed CPT-invariance permits one to express
the probabilities of the inverse decays through the direct ones
\bea
\Gamma_{\bar q \bar q \rightarrow \bar X } = (1+\Delta_q)\Gamma_q,
\,\,\,
\Gamma_{\bar q l \rightarrow \bar X } = (1-\Delta_l)\Gamma_l,
 \nonumber \\
\Gamma_{ q  q \rightarrow  X } = (1-\Delta_q)\Gamma_q,
\,\,\,
\Gamma_{ q \bar l \rightarrow  X } = (1+\Delta_l)\Gamma_l.
\label{inv}
\eea
So direct and inverse decays produce the {\it same} sign of baryon
asymmetry and not the opposite one as is necessary for compensation.
On the other hand we know that in thermal equilibrium no asymmetry is
generated and the excess of baryons produced in the decays is erased by
some other processes. These processes are B-nonconserving $2\rightarrow
2$-scattering of quarks and leptons with X-boson exchange \cite{ad4}.

The mechanism outlined here would be operative if a nonequilibrium number
density of X-bosons was created. Usually massive particles are in
equilibrium at high temperatures, $T \gg m$, and their number density
exceeds the equilibrium one when $T$ becomes comparable to $m$. If they
are unstable, they sooner or later come back to equilibrium because
the decay rate $\Gamma$ remains constant while the expansion rate goes
down. However their number density at this stage is Boltzmann suppressed,
$n\sim \exp (-m/T)$, and is negligibly small. Therefore the most favorable
period for the generation of asymmetry is when $m/T = O(1)$.
If $X$ are gauge (or Higgs) bosons of grand unification the situation is
somewhat more complicated because they might never be in equilibrium at
early stage of the universe evolution. First, even if the temperature of
the primeval plasma was higher than the grand unification scale, $m_{GUT} =
10^{16} -10^{15}$ GeV, the rate of their production would be smaller than
that of the expansion and the number density of these bosons could always be
smaller than the equilibrium one. To see it one needs to compare the
Hubble parameter given by eq. (\ref{h}) to the decay or two-body reaction
rates. The bosons might be out of equilibrium both in unbroken symmetry
phase (when $m_X = 0$, except for temperature corrections) and in the
broken one. Second, the universe temperature
could always be smaller than $m_{GUT}$ and correspondingly thermally
produced X-bosons would
never have been abundant and their role in the baryogenesis would be
negligible. Another possibility is that the inflaton field is
predominantly coupled to these bosons and though the temperature of the
universe after thermalization was much smaller than their mass, they
might be abundantly produced by the inflaton decay so that their initial
number density was well above the equilibrium one \cite{dl}.
In such a case the decays of
the intermediate bosons of grand unification could produce the baryon
asymmetry of the universe.

If we consider particles of smaller mass, the deviation from equilibrium
is generically rather small (see eq.((\ref{dff})). One possible way to
break the equilibrium is to assume a weaker-than-gauge coupling. Another
possibility is a late generation of mass \cite{lps,yam}, when particles
acquire their mass as a result of phase transition with $T<m$, while their
number density remains those of massless particles, $n\sim T^3$, and is
not Boltzmann suppressed.

There is one more problem usually associated with the GUT-scale
baryogenesis, namely the problem of relic gravitino \cite{kl,ekn}.
The gravitino is a spin-(3/2) particle with the interaction strength
inversely proportional to the Planck mass which appears in supergravity
theories as a superpartner of graviton. The cross-section of their
production/annihilation is roughly $\sigma_{3/2} \approx 1/m_{Pl}^2$.
Correspondingly their number density relative to entropy density
should be equal to
\beq
{n_{3/2}  / s} \approx 10^{-2} {T_{reh} / m_{Pl}} ,
\label{n32}
\eeq
where $T_{reh}$ is the temperature of the reheated universe after
inflation. The decay width of gravitino is
\beq
\Gamma_{3/2} = {m_{3/2}^3 / m_{Pl}^2}
\approx (10^5 {\rm sec})^{-1} \left( {m_{3/2} / {\rm TeV} }\right)^3
\label{gam32}
\eeq
If $T_{reh} \sim T_{GUT}$, gravitini can be abundant at nucleosynthesis
and destroy the good agreement of the theory with observations.
However if the initial state produced by the decay of the inflaton was
considerably out of equilibrium with abundant X-bosons but with the
reheating temperature much below the GUT-scale, the gravitini would not
be created in a dangerous amount.

To conclude, it is probably too early to abandon GUT-baryogenesis.
It naturally has the necessary features: baryonic charge nonconservation
(compatible with the observed proton stability and absence of
neutron-antineutron oscillations),
deviation from thermal equilibrium and may have a sufficiently strong
CP-violation.

\section{Other Models.}

There are quite many scenarios of baryogenesis which are much more
sophisticated
than the simple ones based on heavy particle decays. They all can take
place at much smaller temperatures or energies than $T_{GUT}$. This
energy range may even be  accessible in the acting accelerators.
The (possibly incomplete) list of scenarios includes:
\begin{enumerate}
\item{}
Electroweak baryogenesis.
\item{}
Baryogenesis by supersymmetric baryonic charge condensate.
\item{}
Spontaneous baryogenesis.
\item{}
Baryo-through-lepto-genesis.
\item{}
Baryogenesis in black hole evaporation.
\item{}
Baryogenesis by topological defects (domain walls, cosmic strings,
magnetic monopoles).
\end{enumerate}

Below we will briefly discuss only the first three ones. A more detailed
discussion of them, as well as of the remaining,  and a list of
references can be found in the review paper \cite{ad1}.
\bigskip

{\it 7.1.  Electroweak baryogenesis \cite{krs}. }

\bigskip

This model is the most fashionable
now and a majority of the recent papers on the subject deals with different
versions of the electroweak scenario. Surprisingly the standard model of
electroweak interactions have all the necessary ingredients for
successful baryogenesis. It is known from experiment that C and CP symmetries,
are broken. Theoretically introduction
of CP-nonconservation into the standard model is easily done either by a
complex quark mass matrix \cite{km} with at least three generations
or, what is essentially the same, by complex coupling constants of
the Higgs fields. What is more surprising is that baryons are not
conserved by the usual electroweak interactions~\cite{gh}. This is a
rather complicated phenomenon and is related to the so called quantum
chiral anomaly \cite{adler,beja}. The classical electroweak Lagrangian
conserves baryonic charge. Quarks always enter in bilinear combinations
$\bar q q$,  so that a quark can disappear only in collision with an
antiquark. In other words the classical baryonic current is conserved:
\beq
\partial_\mu J_\mu^B = \sum_j \partial_\mu (\bar q_j \gamma_\mu q_j) =0.
\label{djcl}
\eeq
However quantum corrections destroy this conservation law and instead of
zero in the r.h.s. one gets
\beq
\partial_\mu J_\mu^B = {g^2 C\over 16\pi^2}  G_{\mu\nu}
\tilde G_{\mu\nu} ,
\label{djqu}
\eeq
where $C$ is a numerical constant,
 $\tilde G_{\mu\nu} =G_{\alpha\beta} \epsilon_{\mu\nu\alpha\beta}/2$,
and the gauge field strength, $G_{\mu\nu}$, is given by the expression
\beq
G_{\mu\nu} = \partial_\mu A_\nu - \partial_\nu A_\mu +
g [A_\mu A_\nu ] .
\label{gmunu}
\eeq
An important fact is that the anomalous current nonconservation is
proportional to the total derivative of a vector operator:
$G_{\mu\nu} \tilde G_{\mu\nu} = \partial_\mu K_\mu$ where the anomalous
current $K_\mu$ is
\beq
K_\mu = 2 \epsilon_{\mu\nu\alpha\beta} \left( A_\nu \partial_\alpha
A_\beta + {2\over 3} ig A_\nu A_\alpha A_\beta \right).
\label{kmu}
\eeq
The last term in this expression does not vanish only for non-Abelian gauge
theories because the antisymmetric product of three vector
potentials $A_\nu$ can be nonzero due to different group indices (e.g.
for the electroweak group it should contain the product of $W^+$, $W^-$ and
the isospin one part of $Z^0$).

Usually a total derivative is unobservable because one can get rid of it
integrating by parts. However, this may be not true for $K_\mu$ (\ref{kmu}).
Indeed the
gauge field strength $G_{\mu\nu}$ should vanish at infinity but the
potential $A_\mu$ does not necessarily vanish. Thus different vacuum states,
which all have $G_{\mu\nu} =0$, differ by the value of $K_0$
(in fact by the integral from that over the space volume). Since the
difference $J_\mu^B - K_\mu $ is conserved, transition from one
vacuum state to another leads to a change in baryonic charge. The path
from one vacuum to another is separated by a potential barrier where
$G_{\mu\nu} \neq 0$. As we know from quantum mechanics the barrier
penetration at small energies is exponentially suppressed, that is why
the probability of processes with $\Delta B \neq 0$ contains the very
small factor, $\exp (-16\pi^2 /g^2 ) \approx 10^{-160}$. However at
high energies or temperatures (comparable or above the barrier height)
the transition between different vacua can be achieved by a classical
motion over the barrier. The hight of the barrier,
as calculated in ref. \cite{sphal},  is around several TeV. In fact
the barrier disappears at high temperatures together with the W or Z
masses according to the law: $m^2_W (T) = m^2_W (0)(1 - T^2/T_c^2) $. This
also occurs in the same TeV region. So one may expect that at high
temperatures baryon nonconservation is not suppressed. It has been argued that
above the electroweak phase transition the processes with $\Delta B \neq
0$ are much faster than the universe expansion, so that any preexisting
baryon asymmetry would be washed out. To be more precise electroweak
interactions (even with the chiral anomaly) conserve the difference
between baryonic and leptonic charges, $(B-L)$. Thus at high temperatures
only $(B+L)$ may be erased while a preexisting $(B-L)$ is conserved.

If this picture is correct, electroweak interactions at high temperatures
play the role of a terminator of asymmetry and not of a creator. If the
electroweak phase transition is second order, then everything goes
smoothly, thermal equilibrium is not disturbed, and asymmetry is not
generated even below the phase transition. The type of the phase
transition depends upon the mass of the Higgs boson (or bosons in
extended models). For high mass the transition is second order while for
a low mass it is first order. The boundary value of the mass is not well
known even in the minimal standard model and different estimates give
the values
somewhere between 50 and 100 GeV. If the Higgs mass is below this boundary
value, then phase transition is first order and the deviation from
thermal equilibrium can be significant. In this case there would be both
phases coexisting in the primeval plasma. In the symmetric (high temperature)
phase the processes with $\Delta (B+L) \neq 0$ are fast and the asymmetry
is washed out. In the broken symmetry (low temperature) phase everything
is conserved and the asymmetry remains constant. Thus baryogenesis
could proceed only in the boundary region between the phases.

We see that the
standard model in principle has all the necessary properties for creation of
the baryon asymmetry. However,
now a consensus is reached that in the minimal standard model the effect
is by far too small to create the observed asymmetry, because CP-violation
is extremely weak. There are  chances for success in extended models
with several Higgs fields like e.g. low energy SUSY. Still
the asymmetry could be washed out even in the broken phase because the
preexponential factor can be quite large \cite{rs,dd}.

However, all of the above may be wrong by the following reason. Processes of
quark and lepton transformation with a nonzero change of baryonic (and
leptonic) charge at high temperatures are accompanied by a change in the
structure of the gauge and Higgs fields. Roughly speaking the classical
field configuration should be present in the course of the transition,
the so called sphaleron \cite{sphal}:
\begin{eqnarray}
A^{sph}_k = {i\epsilon_{klm} x^l\tau^m \over r^2} f_A(\xi), \nonumber \\
\phi^{sph} = {i\eta \over \sqrt 2} {\tau^i x_i \over r}
\left(0,\,1 \right) f_\phi (\xi),
\label{sphal}
\end{eqnarray}
where $\xi = g\eta r$, $\eta$ is the vacuum expectation value of the Higgs
field, and the functions $f$ have the properties $f(0) =1$ and
$f(\infty)=0$. The size of this object is much larger than its Compton wave
length, that is why it is called a classical field configuration.
It is assumed that sphalerons are in thermal equilibrium,
so that their number density is determined by the Boltzmann exponent,
$\exp(-F/T)$, where $F$ is its free energy. In the broken symmetry phase $F
=O({\rm TeV})$, while in the symmetric phase $F\sim T$. If this is true, the
processes with baryon nonconservation are not suppressed at high
temperature. However the rate of production of classical field states in
the collision of elementary particles is not known and we cannot say if
they are in equilibrium or not. An analogy with magnetic monopoles in
non-Abelian gauge theories \cite{amp,th1} (which are also classical states),
shows that production of such states in two-body or few-body
collision is exponentially suppressed. Nothing is known about the
probability of production of monopoles or sphalerons
in, say, a hundred-particle collision.
Presumably to create a pair of monopole-antimonopole or a sphaleron, one
needs to create a special coherent field configuration which is almost
improbable in the primeval plasma. If this is true then electroweak
processes do not produce or destroy baryons in significant amount.
No analytical way to solve this problem is known at the present stage.
The processes, which should be considered, are essentially non perturbative
and multi-particle ones. The only available approach to the calculation of the
sphaleron transition rates is a numerical lattice simulation. Unfortunately
the results by different groups are conflicting and vary from
zero \cite{asy,pa} to  the non suppressed one~\cite{ak},
$\Gamma \sim (\alpha_w T )^4 $.

%\newpage
\bigskip

{\it 7.2. Baryonic charge condensate \cite{afd}.}

\bigskip

Supersymmetric theories opened new possibilities for baryogenesis, first,
because in many of them baryonic charge is not conserved even at the
energies below GUT scale and, second, because there exist scalar fields
with nonzero baryonic charge: superpartners of quarks (they are denoted
$\chi$ in what follows). The potential for
these fields may have the so called flat directions along which the field
can evolve without changing energy or, in other words, the field in
this direction is massless. A massless scalar field is known to
be infrared unstable in the de Sitter background \cite{bd} and its quantum
fluctuations rise as $\langle \phi^2_0 \rangle = H^3 t /(2\pi)^2$ where $H$
is the Hubble parameter describing the universe expansion, $a(t) \sim
\exp (Ht)$. For nonzero mass, but in the case that $H\gg m$, the rise of
the field is stopped when its potential energy becomes equal to the
kinetic one, $U(\phi) \sim H^4$. The calculations \cite{phim,vifo} give
$\langle \phi^2_m \rangle = 3H^4 /8\pi^2m^2$. The wave lengths of quantum
fluctuations are exponentially stretched up together with the expansion and
during inflationary stage "classical" condensates of light scalar fields
can be developed. This condensate may store baryonic charge (if the
field, like e.g. $\chi$, possesses it) and when inflation is over, the
decay of $\chi$ would produce a nonzero baryon asymmetry. The picture is
slightly more complicated by the following reasons. First, the field
$\chi$ should not possess any conserved quantum numbers. The current
conservation condition
\beq
D_\mu j^\mu = \partial_\mu j^\mu + 3H j_0 = 0
\label{dmuj}
\eeq
results in vanishing of any conserved current density as $j_0 \sim \exp
(-3Ht)$. So the interesting candidates for baryogenesis are the fields
which are electrically neutral and colorless.  Second, the contribution
of $\chi$ into baryonic current is given by the expression:
\beq
j^{(\chi)}_\mu = i\left[\chi^* \partial_\mu \chi-
\left( \partial_\mu \chi^* \right) \chi \right]
= -2 |\chi|^2 \partial_\mu \theta ,
\label{jchi}
\eeq
where $\theta$ is the phase of the field $\chi$. The baryonic charge
density can be visualized as the angular momentum of the
mechanical rotation of a point-like body in the two-dimensional plane
$({\rm Re} \chi, {\rm Im}\chi)$ in the potential field $U(\chi)$. If this
potential is spherically symmetric, i.e. $U=U(|\chi|)$, then the
angular momentum or, in other words, baryonic charge is conserved. A
potential which does not conserve $B$ and has flat directions for
$m=0$ and $|\lambda_2| = \lambda_1$  can e.g. be written as
\beq
U(\chi) = m^2 |\chi|^2 + {1\over 2} \lambda_1 |\chi |^4
+ {1\over 4} \lambda_2 \left( \chi^{*4} + \chi^4 \right) .
\label{uchi}
\eeq
In this toy model the field $\chi$ would evolve as follows. During
inflationary stage it would travel along the valley far from the origin.
When inflation is over the field would relax down to $\chi =0$. If
the phase of the field tended to a constant value during inflation,
so that $\dot \theta =0$ and, moreover,
the field itself was exactly on the bottom of the
valley, then it would come down to origin without rotation i.e. with zero
baryonic charge. No asymmetry would be generated in this case. However
because of quantum fluctuations during inflation (or, what is the same,
by production of quanta of the field $\chi$ by the background curved
space-time) there should be some
motion and displacement orthogonal to the direction of the valley.
The energies of such motion and displacement are
$|\chi|^2 (\partial_\mu \theta)^2 \sim \lambda |\chi |^4  \sim H^4$.
These angular excitations would
give rise to a rotation when $| \chi |$ evolves down to zero to the region
where the potential is approximately spherically symmetric. The
subsequent decay of $\chi$ into fermions may produce quite a
large baryon asymmetry of the universe even if the decay goes with
conservation of baryonic charge. The asymmetry may be close to one.
While in other scenarios the asymmetry is generically smaller than the
observed value and attempts should be taken to get it as large as
possible, here it is another way around: the asymmetry is too large and
one should invent a way to make it smaller. A strong
suppression of the angular motion and respectively of the asymmetry can
be achieved through particle production by the oscillation of $\chi$ in
the direction orthogonal to the valley \cite{dk}.

An unusual feature of this model is that baryon asymmetry can be generated
without explicit C and CP breaking in the Lagrangian. The sign of the
asymmetry is determined by the direction of the rotation of $\chi$ and
the latter is chaotically distributed over the initial $\chi$
configuration created after inflation. It means that in such a scenario
the universe as a whole is charge symmetric and there should be equally
abundant domains of matter and antimatter (see below, Sec. 8).
The characteristic size of these domains is about \cite{ad1}
\beq
l_B \approx H^{-1}_I \exp \left( -1/\sqrt \lambda \right).
\label{lbad}
\eeq

In another version of this model \cite{rand} an explicit CP-violation
has been introduced, so that the direction of the valley has been shifted
away from the center. Due to this misplacement the field
$\chi$ comes down to the origin with already a nonzero angular momentum
and the asymmetry has a definite sign determined
by the direction of the valley.

\bigskip

{\it 7.3. Spontaneous baryogenesis \cite{ck2}.}

\bigskip

If a theory possesses a $U(1)$-symmetry which is either generated by
baryonic charge or by a charge which is not orthogonal to the baryonic
one, then spontaneous breaking of this symmetry results in nonconservation
of baryonic charge of physical particles (as we see below and is well
known, the total charge, i.e. the charge of particles plus vacuum is
conserved). As a result a baryon asymmetry
may be generated in the broken symmetry phase.
This kind of phenomenon may take place in some
electroweak scenarios with several Higgs fields where their relative
phase plays the role of Goldstone boson which appears after a spontaneous
symmetry breaking.

Generically the model of spontaneous breaking of a global symmetry is
described by a scalar field $\phi$ with the potential
\beq
U(\phi) = \lambda (|\phi|^2 - \eta^2)^2 ,
\label{uphi}
\eeq
where $\eta$ is a constant c-number. In the lowest
energy state in this potential (vacuum) the field $\phi$ is non-vanishing,
$\phi = \eta \exp (i\theta)$. A particular choice of the vacuum state
among many degenerate ones, corresponding to different values of $\theta$,
results in the spontaneous symmetry breaking. The
field $\theta (x)$ is called the Goldstone boson. If there is no explicit
symmetry breaking but only the spontaneous one, the theory is invariant
with respect to the transformation
\beq
\theta(x) \rightarrow \theta (x) + const
\label{thetac}
\eeq
This means that the field $\theta$ is massless. In other words the curve
where the potential $U(\phi)$ (\ref{uphi}) reaches its
minimum, is flat and $\theta$ can
evolve along  this curve without changing energy.
If the bottom of the potential
is tilted, so that the degeneracy in the potential energy of $\theta$
disappears, we speak about an explicit symmetry breaking (as e.g. in the
axion case). In this case the $\theta$-field typically acquires a nonzero
mass and becomes a pseudo-goldstone boson.

Let us consider the following toy model with the scalar field $\phi$ and
two fermionic fields "quarks" $Q$ and "leptons" $L$. The theory is
supposed to be invariant with respect to the "baryonic" $U(1)$-symmetry:
$\phi \rightarrow \exp(i\alpha) \phi$,
$Q \rightarrow \exp(i\alpha) Q$, and $L \rightarrow L$, where $\alpha$ is
a constant phase. The corresponding Lagrangian has the form:
\beq
{\cal L} = (\partial \phi )^2 - U(\phi) +
i\bar Q \gamma_\mu \partial_\mu Q + i\bar L \gamma_\mu \partial_\mu L +
(g \phi \bar Q l + h.c.).
\label{lphi}
\eeq
where $U(\phi)$ is given by eq. (\ref{uphi}) and h.c. means hermitian
conjugate. In the spontaneously broken
phase when $\phi = \eta \exp (i\theta)$, the Lagrangian can be rewritten as:
\beq
{\cal L} =\eta^2 (\partial \theta )^2 - V(\theta) +
i\bar Q \gamma_\mu \partial_\mu Q + i\bar L \gamma_\mu \partial_\mu L +
\left[ g \eta \exp (i\theta) \bar Q L + h.c.\right]+...
\label{ltheta1}
\eeq
where the potential $V(\theta)$ describes a possible explicit symmetry
breaking, which is not present in the original Lagrangian (\ref{lphi}),
and the radial degrees of freedom are supposed to be very heavy and are
neglected.

Another representation of this Lagrangian may be useful, namely if we
introduce the new quark field by rotation $Q \rightarrow \exp(i\theta) Q$,
then we get
\beq
{\cal L} =\eta^2 (\partial \theta )^2 + \partial_\mu \theta j^B_\mu
- V(\theta) +
i\bar Q \gamma_\mu \partial_\mu Q + i\bar L \gamma_\mu \partial_\mu L +
(g \eta \bar Q L + h.c.),
\label{ltheta2}
\eeq
where $j_\mu^B = \bar Q \gamma_\mu Q$ is the baryonic currents of quarks.
In this expression the interaction of $\theta$ with matter fields enters
only linearly. It is imperative that the current $j^B_\mu$ is not
conserved, otherwise the interaction term
\beq
{\cal L}_{int}  = \partial_\mu \theta \,j^B_\mu
\label{lint}
\eeq
can be integrated away. This current is indeed nonconserved.
Combining the equations of motion for $Q$ and $L$ one sees that
$\partial_\mu j_\mu^B = ig\eta (\bar L Q - \bar Q L  )$.

For the case of a homogeneous, only time-dependent field $\theta$,
the expression (\ref{lint}) can be written as $ \dot \theta n_B$ where $n_B$
is the density of baryonic charge. Therefore one is tempted to identify
$\dot \theta$ with the baryonic chemical potential \cite{ck2,ck1}. If this
were so the baryonic charge density would be nonzero even in thermal
equilibrium, when the reaction rates are fast, while $\theta$ is not
relaxed down to the dynamical equilibrium point at the minimum of the
potential, where $\dot \theta =0$. The
charge density for small $\dot \theta$ would be equal to
\beq
n_B = B_Q \dot \theta T^2 /6,
\label{nbt}
\eeq
where $B_Q$
is the baryonic charge of the quarks $Q$. It is not the case, however, as
can be seen immediately from the equation of motion for the
$\theta$-field \cite{df}:
\beq
2 \eta^2 \partial^2 \theta = -\partial_\mu j_\mu ^B
\label{d2theta}
\eeq
In fact this equation is just the law of the total current conservation,
$\partial_\mu J_\mu ^{tot} =0$, where $J_\mu ^{tot}$ is the total baryonic
current including the contribution from the scalar field $\phi$. Though
the symmetry is spontaneously broken the theory still "remembers" that it
was symmetric. In the case of space-point independent $\theta =\theta (t)$ the
equation (\ref{d2theta}) is reduced to $2\eta^2 \ddot \theta = -\dot n_B$.
It can be easily integrated giving:
\beq
\Delta n_B = - \eta^2 \Delta \dot \theta
\label{deltanb}
\eeq
which is evidently incompatible with eq. (\ref{nbt}). One should
definitely trust eq. (\ref{deltanb}) because this is simply the
condition of total current conservation which is not disturbed
by thermal corrections. Below we will discuss in
some detail why $\dot\theta$ cannot be interpreted as the baryonic chemical
potential, and  thus why eq. (\ref{nbt}) is incorrect,
but let us first consider the generation of baryon asymmetry
both in the pure goldstone and pseudo-goldstone cases.
We have just seen that in the goldstone case the baryonic charge
density is given by eq. (\ref{deltanb}). The initial value of
$\dot\theta$ is determined by inflation and depends on whether the symmetry
was broken prior to the end of inflation or after that. We assume that the
former is true, then the kinetic energy of the $\theta$-field is
given by \cite{h4} $\eta^2 (\partial \theta)^2 \sim H_I^4$. So
$\dot \theta \sim H_I^2/\eta$ where the Hubble parameter during
inflation, $H_I$, can be found by matching the energy of the
inflaton $\rho_{inf} \sim H^2_I m^2_{Pl}$ and the thermal energy after
reheating $\rho_{reh} \sim T^4_{reh}$. Comparing these expressions we
find that
\beq
\beta \sim {n_B \over T^3} \approx {\eta T_{reh} \over m_{Pl}^2} .
\label{betgold}
\eeq
If the scale of the symmetry breaking, $\eta$,
and the reheating temperature are
not far from the Planck scale the asymmetry would be large enough to
explain the observed value, $\beta \approx 3\cdot 10^{-10}$. However a
serious problem emerges in this scenario. It is known that all regular
classical motions during inflation are exponentially red-shifted down to
zero. The initial non-vanishing $\dot \theta$ came from quantum
fluctuations at the inflationary stage. The characteristic size of the
region with a definite sign of $\dot \theta$ is microscopically small,
$l_B^{inf} \sim H^{-1}$, and even after the red-shift
$z_{reh}+1 = T_{reh}/ 3^o
{\rm K} $ it remains much smaller than the size of baryonic domains now,
$l_B > 10 $ Mpc.

Let us turn now to the pseudo-goldstone case when $\theta$ has a nonzero
potential $V(\theta) = \Lambda^4 \cos \theta$. If $\theta$ is close to
the minimum of this potential it can be approximated by the mass term,
$V(\theta) \approx -1+m^2 \eta^2 (\theta-\pi)^2 /2\,$ with $m^2
=\Lambda^4/\eta^2$. The equation of motion for
$\theta$ now acquires an extra term related to the potential force:
\beq
\eta^2 \ddot \theta +3H\dot \theta + V' (\theta) =
\partial_\mu j_\mu^B.
\label{ddottheta}
\eeq
We have also taken into account the Hubble friction term connected
to the expansion of the universe. We assume that initially $\theta$
is away from
its equilibrium value at $\theta_{eq} = \pi$. It is natural to assume that
$\theta$ can be found anywhere in the interval $(0,2\pi)$ with equal
probability. During inflation when $H\gg m$ the
magnitude of $\theta$ remains
practically constant due to the large friction term, $3H\dot \theta$. The
region with a constant $\theta$ is exponentially inflated,
$l_B \sim l_i \exp (H t)$ and may be large enough to be bigger than
the lower limit to the size of baryonic domain today. When inflation is
over and the Hubble parameter falls below $m$ we can neglect the Hubble
friction and the field $\theta$ starts to oscillate in accordance with the
equation
\beq
\ddot \theta + m^2 \theta = -\partial_\mu j_\mu^B / \eta^2 .
\label{ddotm}
\eeq
The oscillating $\theta$ would produce both baryons and antibaryons but
with different number densities because the current $j_\mu^B$ is not
conserved.
To calculate the asymmetry in this case the following arguments have been
used in the literature. The equation of motion for $\theta$ with the back
reaction of the produced particles was assumed to be
\beq
\ddot \theta + m^2 \theta + \Gamma \dot \theta =0
\label{ddotgam}
\eeq
This equation has a solution correctly describing the decrease of the
amplitude of $\theta$ due to production of particles, namely
\beq
\theta = \theta_i \exp (-\Gamma t/2) \cos(mt +\delta)
\label{theta}
\eeq
Comparing eqs. (\ref{ddotm}) and (\ref{ddotgam}) one may conclude that
\beq
\partial_\mu j_\mu^B = \eta^2 \Gamma \dot \theta .
\label{djb}
\eeq
However this identification is not correct \cite{df}. It can easily be seen,
that if eq. (\ref{djb}) was true, then the energy of the produced
particles would be larger
than the energy of the parent field $\theta$. This of course cannot be
true. Indeed if the expression (\ref{djb}) was correct then the energy
density of the produced baryons could be estimated as follows. The energy
of each quark produced by the field oscillating with the frequency $m$
is equal to $m/2$. The total number density of the produced quarks,
$n_Q+n_{\bar Q}$, is larger than the density of the baryonic charge, $n_B
= n_Q -n_{\bar Q}$. So the energy density of the produced baryons is larger
than $m n_B $. From eq. (\ref{djb}) follows that $n_B$ is linear in
$\theta$ while the energy density of the field $\theta$ is quadratic in
$\theta$. Thus in the limit of small $\theta$ the energy of the produced
particles would be bigger than the energy of the field-creator. This is
of course impossible and it proves that the identification made above is
wrong. In fact the correct solution of the equation does not necessarily
mean that the equation itself is correct. For example one can describe
the decaying field by the equation
\beq
\ddot \theta + (m-i\Gamma/2)^2 \theta = 0 .
\label{mgam}
\eeq
This equation has the same solution (\ref{theta}) but does not permit to
make the identification (\ref{djb}).

In the paper \cite{df} we have derived in one loop approximation the
equation of motion for
$\theta$ with the account of the back reaction of the produced fermions.
It is a nonlocal nonlinear equation which in the limit of a small
amplitude of $\theta$ has the same solution as equations (\ref{ddotgam})
and (\ref{mgam}) but does not permit to make a wrong identification
(\ref{djb}). The direct calculation of the particle production by the
time-dependent field (\ref{theta}) gives the result \cite{dfrs}
\beq
n_B \sim \eta^2 \Gamma_{\Delta B} (\Delta \theta)^3 ,
\label{theta3}
\eeq
where $ \Gamma$ is the width of $\theta$-decay with
nonconservation of baryonic charge and $\Delta \theta$ is the difference
between the initial and final values of $\theta$. The asymmetry in this
case can roughly be estimated as
\beq
\beta = g^2 (\Delta \theta)^3 {\eta^2 m \over T^3} .
\label{bet3}
\eeq
The size $l_B$ in this scenario depends upon the model parameters and can
be either larger than the present-day horizon or much smaller, inside our
visibility.

Let us turn now to the possibility of the interpretation of $\dot \theta$ as
the baryonic chemical potential. It enters the Lagrangian as
${\cal L}_\theta = \dot \theta n_B$, in exactly the same way as
a chemical potential should enter the Hamiltonian. However from the relation
between ${\cal L}$ and ${\cal H}$
\beq
{\cal H} = {\partial {\cal L} \over \dot \phi} \dot \phi - {\cal L},
\label{hl}
\eeq
follows that the contribution from ${\cal L}_\theta$ into
the Hamiltonian formally
vanishes. The Hamiltonian depends upon $n_B$ through the canonical momentum,
$P= \partial {\cal L} / \partial \dot \theta = 2\eta^2 \dot \theta + n_B$. So
from the kinetic term in the Lagrangian, $\eta^2 (\partial \theta )^2$, one
gets
${\cal H} = (P - n_B)^2/4\eta^2 $. If however the field theta is an external
one (let us denote it now as capital $\Theta$), so that
the Lagrangian does not contain its kinetic term, and $\Theta$
only comes there as $\dot \Theta n_B$, then
we do not have any equation of motion for $\Theta$, it is an external
"constant"
variable. In this case the Hamiltonian would be ${\cal H} = - \dot \Theta n_B$
and this $\dot \Theta$ is the baryonic chemical potential. In this case for
sufficiently fast reactions the baryonic charge density would be given by
expression (\ref{nbt}).

However for our dynamical field $\theta$ the equation of motion, which
governs its behavior does not permit $\theta$ to be an adiabatic variable
which can change slowly with respect to the reactions with
$\Delta B \neq 0$. The change of baryonic charge implies the similar change
in $\theta$ so equilibrium is never reached. In the pure Goldstone situation
this is seen of course from the equation of motion (\ref{d2theta}).
For the pseudo-goldstone
case the situation is slightly more complicated but still the result is the
same. Let us consider the Dirac equation for quarks in the presence of
theta-field:
\beq
\left( i \gamma_\mu \partial_\mu - \dot \theta \right) Q = -g\eta L
\label{dirq}
\eeq
We neglected here a possible mass term which is not essential.
In perturbation theory one is tempted to neglect the r.h.s. of this equation
because it is proportional to the small coupling constant $g$ and
to study the spectrum of the Dirac equation with zero r.h.s.
The dispersion relation for this
equation is $E = p \pm \dot \theta$ where signs "+" and "-" stand respectively
for quarks and antiquarks. Thus energy levels
of particles and antiparticles are shifted
by $2\dot \theta$ and in equilibrium their number densities should be
different.
The point is, however, that the change in the population numbers proceeds with
the same speed as the change in $\theta$ or, in other words, the current
nonconservation which can create a difference between $Q$ and $\bar Q$ is
proportional to the same coupling constant $g$ which enters the
equation of motion (\ref{d2theta}) and governs the behavior of
$\theta(t)$ in the Goldstone case. In the pseudo-goldstone case the variation
of $\theta$ can be dominated by the potential term (\ref{ddottheta}).
Hence it may change (oscillate) faster than just in the
limit of zero potential (Goldstone limit)
and one has even less ground to suppose that $\theta (t)$ is an
adiabatic variable. In this case the situation is  worse than in the
Goldstone case because the rate of
variation of baryonic charge is much slower than the variation of $\theta$
and the system is even further from equilibrium.

It may be instructive to see how different fermion/antifermion levels are
populated in the presence of the theta-field in the "rotated" fermion
representation, $Q \rightarrow \exp ( i\theta) Q $,
when the Dirac equation has the form
\beq
 i \gamma_\mu \partial_\mu Q = -g \eta L \exp \left( -i\theta \right) .
\label{dirq1}
\eeq
This equation, in the limit of $g=0$, has the same spectrum for particles and
antiparticles, $E=p$, but the levels would be differently populated because
the interaction term (in the r.h.s.) does not conserve energy.
Assuming that $\theta (t)$ is a slowly varying function of time we can write
$\theta (t) \approx \dot \theta t$. Thus in the reactions with quarks their
energy is increased by $\dot \theta$ in comparison with the energy of the
participating particles, while the energy of antiquarks would decrease by the
same amount. One sees from this example that the energies of particles and
antiparticles are indeed getting different but the process of differentiation
is proportional to the coupling $g$. The arguments presented above, concerning
the possibility of the interpretation of $\dot \theta$ as the baryonic chemical
potential, are based on discussions with K. Freese, R. Rangarajan, and
M. Srednicki.

\section{Antimatter in the Universe.}

The scale $l_B$ which characterizes the size of the domain with the dominance
of
matter is not known, neither from theory nor from observations.
In fact different theoretical models give certain predictions about $l_B$ but
they are strongly parameter dependent and, even worse, we do not know which
model is a true one. In the standard GUT-baryogenesis the magnitude of
baryon asymmetry $\beta$ is constant over all the universe so $l_B$ is either
infinitely large (in an open universe) or is equal to the universe size (in
a closed universe).  In this case there would be no place in the world with
a substantial amount of antimatter. This is not obligatory however, and
many scenarios of baryogenesis do not possess this property. In particular
it is possible to create a universe
which is charge symmetric as a whole with domains of matter
alternating with domains of antimatter. Even in this case it is not excluded
of course that $l_B$, though finite, is larger than the present-day horizon.
If so, one cannot distinguish observationally the two possibilities of a
charge symmetric universe and a completely asymmetric one. However,
if we are lucky,
domains of antimatter may be not so far away and we may have a chance to see
them, in particular by observation of antinuclei in cosmic rays. The
present-day experimental situation and prospects for the future search of
antimatter is reviewed at this School by V.Plyaskin \cite{vpl} (see also
the recent paper \cite{der}).
Theoretical models predicting an abundant amount of
antimatter inside our visibility region are reviewed in refs. \cite{ad5,ad6}.

The general conditions for cosmological creation of both matter and antimatter
with sufficiently small scale parameter $l_B$ are:
\begin{enumerate}
\item{}Different signs of C and CP-violation in different space points.
\item{}Inflationary (but moderate) blow-up of regions with different signs of
charge symmetry breaking.
\end{enumerate}
The first condition can be realized in models with spontaneous breaking
of charge symmetry \cite{lee}. One
can see that domains with opposite signs of C(CP)-odd
phase are indeed formed through this mechanism.
In these domains an excess of either matter or antimatter is
generated by baryogenesis \cite{brst}, depending upon the sign of CP-odd
amplitude.

These models encounter two serious problems. First, the average size of the
domains is too small. If they are formed in a second order
phase transition, their size at the moment of formation is determined by the
so called Ginsburg temperature and is approximately equal to
$l_i=1/(\lambda T_c)$ where $T_c$ is the critical
temperature at which the phase
transition takes place and $\lambda$ is the self-interaction coupling constant.
In this case different domains would expand together with the universe and now
their size would reach $l_0 = l_i (T_c/T_0) = 1/(\lambda T_0)$ where
$T_0 = 2.7$K is the present day temperature of the background radiation. If the
phase transition is first order then the bubbles of the broken phase are
formed in the symmetric background. In this case different bubbles are not
initially in contact with each other, typically the distance between them
is much larger than their size, and their walls may expand faster than
the universe, even as fast as the speed of light. Thus at the moment when the
phase transition is completed the typical size of the bubbles may be as large
as the horizon, $l_f\approx t_f \approx m_{Pl}/T_f^2 $. After that they are
stretched out by the factor $T_f/T_0$ due to the universe expansion.
To make the present day size around (or larger than) 10 Mpc we need
$T_f \sim 100 $ eV. It is
difficult (if possible) to arrange that without distorting successful
results of the standard cosmology. Thus to make
an observationally acceptable size
of the matter-antimatter domains, a super-luminous cosmological expansion seems
necessary. This solution was proposed in ref. \cite{sato} where exponential
(inflationary) expansion was assumed. With this expansion law it is quite easy
to over-fulfill the plan and to inflate the domains beyond the present day
horizon. Effectively it would mean a return to the old charge asymmetric
universe without any visible antimatter. So some fine-tuning is necessary which
would permit to make the domain size above 10 Mpc and below 10 Gpc.
As we have already mentioned the lower bound on $l_B$ presented in
ref.~\cite{der} are much more restrictive. However they may be not applicable
to some particular models (especially with isocurvature fluctuations) and
there still may be some room for antimatter inside the present-day horizon.

The second cosmological problem which may arise in these models is a
very high energy density and/or large
inhomogeneity created by the domain walls~\cite{zko}. This
can be resolved if domain walls were destroyed at a later stage by the symmetry
restoration at low temperature or by some other mechanism \cite{ms,kts}.
However there could be scenarios of baryogenesis in which domains of
matter-antimatter may be created without domain walls. The basic idea of
these scenarios is that baryogenesis proceeds when the (scalar) field which
creates C(CP)-breaking or stores baryonic charge is not in the dynamically
equilibrium state, as it takes place e.g. in the spontaneous
baryogenesis scenario or
in the model of baryogenesis with SUSY baryonic charge condensate (see
Sec. 7). In these
cases charge asymmetry is created by asymmetric initial conditions which
in turn are created by quantum fluctuations during inflationary stage.
In such scenarios one does not need an explicit C or CP violation for
baryogenesis. Moreover domains of matter or antimatter which could be created
in these models do not have any domain walls with a high energy density
so this problem is avoided.

There is quite a rich spectrum of possibilities for objects made of antimatter,
which is open by different models of this kind. There may be just simple
regions like our neighborhood either with matter or antimatter with
sufficiently large sizes. A more curios possibility has been considered
in ref.~\cite{ds}. A mechanism has been proposed there which could create
regions with relatively small sizes and a very high value of the
asymmetry, $\beta = 0.01 - 1$. The sign of the asymmetry with almost
equal probability may be positive or negative. Such regions would mostly
form primordial black holes and, if so, it would be impossible to
distinguish whether
they are formed by baryonic matter or antimatter. But on the
tail of the distribution there might be anti-stars or clouds of antimatter,
enriched by heavier elements (because of large $\beta$ primordial
nucleosynthesis would give larger primordial abundances of heavy elements).
Another exotic possibility is a quasiperiodic
universe filled with alternating baryonic and anti-baryonic layers \cite{dikn}.
For more detail one may address refs. \cite{ad5,ad6}.
These models are of course very speculative and there are neither
theoretical nor experimental arguments in favor of their necessity.
Still they are permitted and, as we know, everything which is not forbidden
has a right to exist. Second, the picture of a charge symmetric universe is
more
attractive than asymmetric one. And last, but not the least,
a search for antimatter will not necessarily  be successful but still
something interesting may be found in the way.

\section{Conclusion.}

We see that cosmology provides really strong arguments in favor of
nonconservation of baryonic charge. Though the Minimal Standard Model (MSM) in
particle physics predicts that baryons are indeed nonconserved, this model
seems to be unable to produce enough baryons for
an explanation of the observed
asymmetry. The necessity for baryogenesis is a strong indication
that there should be a new physics
beyond the Standard Model. We do not know if this is just the low energy
supersymmetric extension of MSM or baryogenesis
demands something new at higher energies.
Possibly the next generation of accelerators will be able to resolve this
very important issue. A very essential for the understanding of the
dynamics of baryogenesis would be an observation of cosmic antimatter.
However such observation (or non-observation) will be able only to
confirm baryosymmetric cosmology but not to reject it.

\section*{Acknowledgments}  This work was supported in part by the Danish
National Research Foundation
through its establishment of the Theoretical Astrophysics Center.
I am grateful to S. Hansen for critical reading the manuscript.

\section*{References}
%\newpage

\end{document}